\documentclass[
aps,
twocolumn,
superscriptaddress,
amsmath,amssymb
]{revtex4-1}

\usepackage{graphicx}
\usepackage{dcolumn}
\usepackage{bm}
\usepackage{color}

\begin{document}

\title{Quantum superposition, entanglement, and state teleportation of a microorganism \\on an electromechanical oscillator}
\author{Tongcang Li}
 \email{tcli@purdue.edu}
 \affiliation{Department of Physics and Astronomy, Purdue University, West Lafayette, IN 47907, USA}
 \affiliation{School of Electrical and Computer Engineering, Purdue University, West Lafayette, IN 47907, USA}
 \affiliation{Birck Nanotechnology Center, Purdue University, West Lafayette, IN 47907, USA}
 \affiliation{Purdue Quantum Center, Purdue University, West Lafayette  IN 47907, USA}
\author{Zhang-Qi Yin}
\email{yinzhangqi@mail.tsinghua.edu.cn}
 \affiliation{Center for Quantum Information, Institute of Interdisciplinary Information Sciences, Tsinghua University, Beijing, 100084, China}

\date{\today}

\begin{abstract}
Schr{\"o}dinger's thought experiment to prepare a cat in a superposition of both alive and dead states reveals profound consequences of quantum mechanics and has attracted enormous interests. Here we propose a straightforward method to create quantum superposition states of a  living microorganism by putting  a small cryopreserved bacterium on top of an electromechanical oscillator. Our proposal is based on  recent developments that the center-of-mass oscillation of a 15-$\mu$m-diameter aluminium membrane has been cooled to its quantum ground state [Nature 475: 359 (2011)], and entangled with a microwave field [Science 342: 710 (2013)].  A microorganism with a mass much smaller than the mass of the electromechanical membrane will not significantly affect the quality factor of the membrane and can be cooled  to the quantum ground state together with the membrane. Quantum superposition and teleportation of its center-of-mass motion state can be realized with the help of superconducting microwave circuits. More importantly, the internal states of a microorganism, such as the electron spin of a glycine radical, can be  entangled with its center-of-mass motion and teleported to a remote microorganism. Our proposal can be realized with   state-of-art technologies. The proposed setup is a quantum-limited magnetic resonance force microscope. Since internal states of an organism contain  information, our proposal also provides a  scheme for teleporting  information or memories between two remote organisms.
\end{abstract}
\maketitle

%
%
%
%
%

\section{Introduction}
In 1935, Erwin Schr{\"o}dinger  \cite{Schrodinger1935} proposed a famous thought experiment to prepare a cat in a superposition of both alive and dead states. He imagined that a cat, a small radioactive source, a Geiger counter, a hammer and a small bottle of poison were sealed in a chamber. If one atom of the radioactive source decays, the counter will trigger a device to release the poison. So the state of the cat will be entangled with the state of the radioactive source. After a certain time, the cat will be in superposition of both alive and dead states. The possibility of an organism to be in a superposition state is counter-intuitive and dramatically reveals the profound consequences of quantum mechanics. Besides their importance in studying foundations of quantum mechanics, quantum superposition and entangled states are  key resources for quantum metrology and quantum information \cite{Wineland2013}. Furthermore, the explanation of how a ``Schr{\"o}dinger's cat'' state collapses is the touchstone of different interpretations of quantum mechanics. While many-worlds interpretations propose that there is no wave function collapse \cite{Everett1957,DeWitt1970}, objective collapse theories \cite{Penrose96,diosi1989,ghirardi1986,Bassi13} propose that the wave function collapse is an intrinsic event that happens spontaneously.

The ``Schr{\"o}dinger's cat'' thought experiment has attracted enormous interest since it was proposed eighty years ago \cite{Wineland2013}. Many great efforts have been made to create large quantum superposition states. Superposition states of photons, electrons, atoms, and some molecules have been realized \cite{Hornberger2012}. Wavelike energy transfer through quantum coherence in photosynthetic systems has been observed \cite{Brixner2005,Engel2007}.
Recent developments in quantum optomechanics \cite{Aspelmeyer2014,Chang2010,Yin2013a,Chan11} and electromechanics \cite{Connell2010,Teufel11,Palomaki2013b,Palomaki2013,Suh2012,Suh2014,Wollman2015,Pirkkalainen2015} provide new opportunities to create even larger superposition states experimentally \cite{Nimmrichter13}.
In 2009,  Romero-Isart  et al.  \cite{Romero10} proposed  to optically trap a living microorganism in a high-finesse optical cavity in vacuum to create a superposition state. That paper increased our hope to experimentally study the quantum nature of living organisms \cite{Schrodinger1935, Schrodinger1944, Bull2015}. Meanwhile, Li et al. demonstrated optical trapping of a pure silica microsphere in air and vacuum \cite{Li10}, and cooled its center-of-mass motion from room temperature to about 1.5 mK in high vacuum \cite{Li11}.
Recently, parametric feedback cooling \cite{Gieseler12} and cavity cooling of pure dielectric nanoparticles \cite{Kiesel13, Asenbaum13,Millen15} were also demonstrated. These are important steps towards quantum ground state cooling of a levitated dielectric particle.
Optical trapping of an organism in vacuum, however, has not been realized experimentally. The main difficulty is that the optical absorption coefficient \cite{Jacques2013} of organisms is   much larger than that of a pure silica particle, which can lead to significant heating  of an optically trapped microorganism in vacuum. Recently, Fisher proposed that  the nuclear spin of phosphorus can serve as the putative quantum memory in brains \cite{Fisher2015}.

\begin{figure}[tbp]
\setlength{\unitlength}{1cm}
\includegraphics[width=8.5cm]{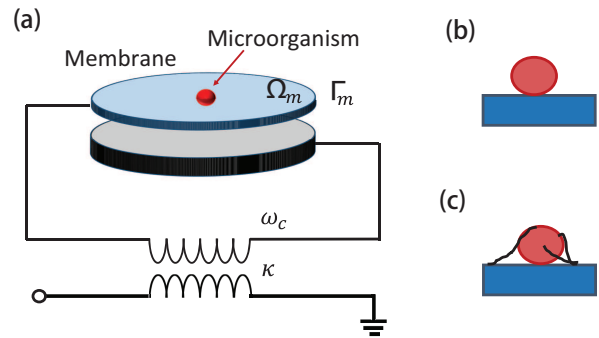}
\caption{(Color online) \textbf{a} Scheme to create quantum superposition states of a microorganism by putting a small bacterium or a virus with a mass of $m$ on top of an electromechanical membrane oscillator with a mass of $M_{\rm mem}$. The membrane is the upper plate of a capacitor embedded in a superconductor inductor-capacitor (LC) resonator. The LC resonator is coupled to a transmission line. The membrane has an intrinsic mechanical oscillation frequency of ${\mathnormal{\Omega}}_{\rm m }$ and a linewidth of ${\mathnormal{\Gamma}}_{\rm m }$ when the microorganism is not  on it. The LC resonator has a resonant frequency of $\omega_c$ and an energy decay rate of $\kappa$. Some microorganisms have smooth surfaces (\textbf{b}) while some have pili on their surfaces (\textbf{c}) }
\label{fig:scheme}
\end{figure}

Here we propose to create quantum superposition and entangled states  of a  living microorganism by putting a small bacterium  on top of an electromechanical oscillator, such as a membrane embedded in a superconducting microwave resonant circuit (Fig. \ref{fig:scheme}). Our proposal also works for viruses. Since many biologists do not consider viruses as living organisms \cite{Bull2015}, we focus on small bacteria in this paper. Our  proposal has several advantages. First, it avoids the laser heating problem as no laser is required in our scheme.  Second, quantum ground state cooling and advanced state control of an electromechanical oscillator integrated in a superconducting circuit have  been realized experimentally \cite{Connell2010,Teufel11,Palomaki2013b,Palomaki2013,Suh2012,Suh2014,Pirkkalainen2015}. Quantum teleportation based on superconducting circuits has also been demonstrated \cite{Steffen2013}. In addition, most microorganisms can survive in the cryogenic environment that is required to achieve ground state cooling of an electromechanical oscillator. Although microorganisms are frozen in a cryogenic environment, they can be still living and  become active after thawing \cite{Mazur1984}. Cryopreservation is a mature technology that has been used clinically worldwide \cite{Mazur1984}. Most microorganisms can be preserved for many years in cryogenic environments \cite{Norman1970}. Even some organs \cite{Fahy2004} can  be preserved at cryogenic temperatures. At millikelvin temperatures, a microorganism can be exposed to ultrahigh vacuum without sublimation of water ice. More importantly, the internal states of a microorganism, such as the electron spin of a glycine radical $\textrm{NH}^+_3{\dot{\textrm{C}}}\textrm{HCOO}^-$ \cite{Hoffmann1995, Zhou2012}, can also be prepared in superposition states and entangled with its center-of-mass motion in the cryogenic environment. This will be remarkably similar to Schr{\"o}dinger's initial thought experiment of entangling the state of an entire organism (``alive" or ``dead" state of a cat) with the state of a microscopic particle (a radioactive atom).

 We also propose to teleport the center-of-mass motion state and internal electron spin state between two remote microorganisms, which is beyond Schr{\"o}dinger's  thought experiment. Since internal states of an organism contain  information, our proposal provides an experimental scheme for teleporting  information or memories between two organisms. Our proposed setup not only can be used to study macroscopic quantum mechanics, but also has applications in quantum-limited magnetic resonance force microscopy (MRFM) for sensitive magnetic resonance imaging (MRI) of biological samples \cite{Degan2009}. This will provide more than  structural information that can be obtained by cryo-electron microscopy \cite{Adrian1984}, which also requires the sample to be cooled to cryogenic temperatures to avoid  sublimation of water ice in vacuum. This system can coherently manipulate and detect the quantum states of electron spins, which enables  single electron spins that could not be read with optical or electrical methods to be used as quantum memory.

\section{The Model}

Recently, a 15-$\mu$m-diameter  aluminum membrane with a thickness of 100 nm has been cooled to quantum ground state by sideband cooling with a superconducting inductor-capacitor (LC) resonator \cite{Teufel11}. Coherent state transfer between the membrane and a traveling microwave field \cite{Palomaki2013}, as well as entangling the motion of the membrane and a microwave field \cite{Palomaki2013b} have been realized.  The experiment in Ref. \cite{Teufel11} was performed in a cryostat at 15 mK. The mechanical oscillation frequency of this membrane is about 10 MHz, with a mechanical quality factor of $3.3 \times 10^5$. The mass of the membrane is 48 pg ($2.9 \times 10^{13}$ Da). As shown in Table \ref{table1}, this mass is about four orders larger than the mass of ultra-small bacteria and even more orders larger than the mass of viruses \cite{Spirin2002,Fuerstenau2001,Ruigrok1984,Luef2015,Zhao2008,Partensky99,Neidhardt96}. We can use this membrane oscillator to create quantum superposition states of a small microorganism. Among different cells, mycoplasma bacteria are particularly suitable for performing this experiment because they are ubiquitous and their sizes are small \cite{Zhao2008}. In addition, they have been preserved at cryogenic temperatures \cite{Norman1970}.  It will also be interesting to create superposition states of prochlorococcus. Prochlorococcus is presumably the most abundant photosynthetic organism on the earth, with an abundance of about $10^5$ cells/mL  in surface seawater \cite{Partensky99, Flombaum13}. It can be used to study quantum processes in photosynthesis.

As shown in Fig. \ref{fig:scheme}a, we consider a bacterium or virus on top of an electromechanical membrane oscillator. We assume the mass of the   microorganism  $m$ is much smaller than the mass of the membrane $M_{\rm mem}$.  Some microorganisms have smooth surfaces (Fig. \ref{fig:scheme}b) while some other microorganisms have pili (hairlike structures) on their surfaces (Fig. \ref{fig:scheme}c). For simplicity, it will be better to use microorganisms with smooth surfaces. One way to put a microorganism on top of an electromechanical membrane is to spray a small amount of microorganisms to air in a sealed glove box, and wait one microorganism to settle down on the electromechanical membrane. The process can be monitored with an optical microscope. After  being cooled down to a cryogenic temperature, the microorganism can be exposed to high vacuum without sublimation of the water ice. Techniques developed in cryo-electron microscopy can be utilized here \cite{Adrian1984}. The sublimation rate of water ice is only  about 0.3 monolayers/h at 128 K with a sublimation energy of 0.45
eV \cite{King2005}.  The sublimation of water ice can be safely ignored at millikelvin temperatures.

At millikelvin temperatures, a frozen microorganism will be in a  hard and brittle state like a glass. It will be stuck on the membrane due to van der Waals force or even stronger chemical bonds.
The pull-off force between a 1 $\mu$m sphere and a flat surface due to van der Waals force is on the order 100~nN \cite{Heim1999}, which is about $10^7$ times larger than the gravitational force on a 1 $\mu$m particle. So the microorganism will move together with the membrane as long as there is a good contact between them.
The oscillation frequency of the membrane oscillator will change by roughly $-{\mathnormal{\Omega}}_{\rm m } m/(2 M_{\rm mem})$, where ${\mathnormal{\Omega}}_{\rm m }$ is the intrinsic oscillation frequency of the membrane. The exact frequency shift  depends on the position of the microorganism on the membrane. Nonetheless, this frequency shift will be  small  and will not significantly affect the ground state cooling or other state control of the membrane when  $m/M_{\rm mem} << 1$.

\renewcommand{\arraystretch}{1.5}
\begin{table}[tbp]
	\centering
		\begin{tabular}{|l|l|l|}
			\hline
			        {\bf Microorganism}          & {\bf Typical mass}  &  $m / M_{mem}$    \\
                                                 &  (pg)  &   ($M_{mem}$ = 48 pg)  \\  \hline
Bacteriophage MS2        &      $6 \times 10^{-6}$      &       $10^{-7}$        \\    \hline
Tobacco mosaic virus &   $7 \times 10^{-5}$     &     $10^{-6}$      \\ \hline
 Influenza virus  &   $3 \times 10^{-4}$     &     $10^{-5}$      \\ \hline
 WWE3-OP11-OD1 &   0.01    &     $10^{-4}$         \\
 ultra-small bacterium &        &              \\ \hline
    Mycoplasma bacterium&   0.02     &     $10^{-4}$         \\ \hline
        Prochlorococcus &   0.3    &     $10^{-2}$         \\ \hline
   E. coli bacterium&    1     &   $10^{-2}$      \\ \hline
		\end{tabular}
	\caption{\label{table1} Comparison of  masses of some viruses and bacteria to the mass of a membrane oscillator ($M_{mem}$ = 48 pg) that has been cooled to the quantum ground state in Ref. \cite{Teufel11}.  These mass values  are obtained from references \cite{Spirin2002,Fuerstenau2001,Ruigrok1984,Luef2015,Zhao2008,Partensky99,Neidhardt96}. }
\end{table}

The change of the quality factor $Q$ of the membrane oscillator due to an attached small microorganism ($m/M_{\rm mem} << 1$) will also be negligible. For a frozen bacterium with a smooth surface (Fig. \ref{fig:scheme}b), the lowest internal vibration frequency can be estimated by the speed of sound in ice divided by half of its size. So the internal vibration modes of the main body will be larger than 1~GHz for a bacterium smaller than 1~$\mu$m. This is much larger than the frequency of the center-of-mass motion of the electromechanical membrane which is about 10 MHz. Thus these internal vibration modes of the main body of a bacterium will not couple  to the center-of-mass vibration of the membrane. The situation will be a little  complex for bacteria that have pili  on their surfaces (Fig. \ref{fig:scheme}c) \cite{Proft2009}. For a very thin and long pilus, it is possible that its vibration frequency to be on the same order or even smaller than the center-of-mass vibration frequency of the electromechanical membrane. However, because the quality factors of both the pili and electromechanical membrane are expected to be very high at milikelvin temperatures, it is unlikely that their frequencies matches within their linewidths to affect the ground state cooling of the center-of-mass mode of the membrane. One can also avoid this problem by embedding the pili in water ice.

\section{Center-of-mass motion: cooling, superposition and teleportation}
We assume the frequency of the center-of-mass motion of  the microorganism and the membrane together to be $\omega_{\rm m }$, which is almost the same as ${\mathnormal{\Omega}}_{\rm m }$.
The motion of the membrane alters the frequency $\omega_c$ of the superconducting LC resonator.
The frequency $\omega_c$ can be approximated with $\omega_c(x) = \omega_0 + G x$, where $x$ is
the displacement of membrane, and $G=\partial \omega_c / \partial x$.
The parametric interaction Hamiltonian  has the form $H_\mathrm{I}=
 \hbar G a^\dagger a \hat{x}= \hbar G\hat{n} x_0 (a_{\rm m } + a_{\rm m }^\dagger)$, where $a$ ($a_{\rm m }$) and $a^\dagger$ ($a_{\rm m }^\dagger$)
are the creation and annihilation operators for LC (mechanical) resonator, $\hat{n}$ is the photon number
operator, and $x_0=\sqrt{\hbar/2M_{\rm mem}\omega_{\rm m }}$ is the zero point fluctuation for mechanical mode.
We denote the single-photon coupling constant $g_0=Gx_0$.

In order to enhance the effective coupling between LC resonator and
mechanical oscillator, the LC resonator is strongly driven with frequency $\omega_{\rm d}$ and Rabi frequency ${\mathnormal{\Omega}}_{\rm d}$.The
corresponding Hamiltonian reads $H_{\rm d}= \frac{{\mathnormal{\Omega}}_{\rm d}}{2} {\rm e}^{-{\rm i}\omega_{\rm d} t} a + \mathrm{h.c.}$. The detuning between the
driving microwave source and the LC resonator is ${\mathnormal{\Delta }}= \omega_{\rm d} -\omega_0$. The steady state amplitude can be
approximated as $\alpha= {\mathnormal{\Omega}}_{\rm d}/(2{\mathnormal{\Delta }} + {\rm i}\kappa)$, where $\kappa$ is the decay rate of the LC resonator. We assume
that the steady state amplitude $\alpha$ is much larger than $1$. We expand
the Hamiltonian with $a-\alpha$, and the total Hamiltonian reads
\begin{equation}\label{eq:Htotal}
H= \hbar {\mathnormal{\Delta }} a^\dagger a + \hbar \omega_{\rm m } a_{\rm m }^\dagger a_{\rm m } + \hbar g(a^\dagger+ a)( a_{\rm m }^\dagger + a_{\rm m }),
\end{equation}
where $g=\alpha g_0$. The detuning can be freely chosen to satisfy the requirements of different applications. In order to cool the
membrane resonator, we choose ${\mathnormal{\Delta }}= -\omega_{\rm m }$. By using rotating wave approximation under the condition
that $\omega_{\rm m } \gg g$, we get the effective equation as \cite{Yin2015}
\begin{equation}\label{eq:Heff}
  H_{\mathrm{eff}} =  \hbar g a^\dagger a_{\rm m } + \hbar g a a_{\rm m }^\dagger.
\end{equation}

In order to cool the membrane oscillation to the ground state,  the system should fulfill the sideband limit  $\omega_{\rm m } \gg \kappa, \gamma_{\rm m }$, where $\gamma_{\rm m }$ is the decay rate of mechanical mode of the whole system \cite{Wilson-Rae2007,Marquardt2007}. $\gamma_{\rm m }$ will be almost the same as the decay rate ${\mathnormal{\Gamma}}_{\rm m }$  of the mechanical mode of the electromechanical membrane alone when the frequencies of internal modes of the microorganism do not match the frequency of the electromechanical membrane.  The master equation for the system is
\begin{equation}\label{eq:master}
  \dot{\rho}= -{\rm i}[H_{\rm eff}, \rho]/\hbar + \mathcal{L}_a \rho + \mathcal{L}_{a_{\rm m }} \rho,
\end{equation}
where the generators are defined by $\mathcal{L}_a \rho =\kappa D_a \rho$,
$\mathcal{L}_{a_{\rm m }} \rho = (1+\bar{n}_{\rm m }) \gamma_{\rm m } D_{a_{\rm m }} \rho + \bar{n}_{\rm m }
\gamma_{\rm m } D_{a_{\rm m }^\dagger} \rho$, and the notation of the generators has Lindblad
form $D_x \rho= 2x \rho x^\dagger -x^\dagger x \rho - \rho x^\dagger x$.
In the limit that $\kappa \gg g$, we can adiabatically eliminate the $a$ mode, and
get the effective master equation
\begin{equation}\label{eq:master1}
  \dot{\rho}=  \mathcal{L}_{a_{\rm m }} \rho + \mathcal{L}_{a_{\rm m }}' \rho,
\end{equation}
where $\mathcal{L}_{a_{\rm m }}' = \kappa' D_{a_{\rm m }}$, and $\kappa' = g^2/\kappa$. We define the total generator
$\mathcal{L}= \mathcal{L}_{a_{\rm m }} + \mathcal{L}_{a_{\rm m }}' = (1+\bar{n}_{\rm m }') \gamma' D_{a_{\rm m }} + \bar{n}_{\rm m }' \gamma' D_{a_{\rm m }^\dagger}$,
where $\gamma'=\gamma + \kappa'$, and $\bar{n}_{\rm m }' = \bar{n}_{\rm m }\gamma/\gamma' $.  As long as $\kappa' > \bar{n}_{\rm m }\gamma$,
the steady state mean phonon number of mechanical resonator $\bar{n}_{\rm m }'$ is less than $1$, which is in the quantum regime.
As a bacterium or virus are attached on the top of membrane, it is also cooled down to the quantum regime.

Once the mechanical mode is cooled down to the quantum regime, we can prepare the mechanical superposition state by
the method of quantum state transfer between mechanical and LC resonators.
For example, we can first generate the superposition state $|\phi_0\rangle= (|0\rangle + |1\rangle)/\sqrt{2}$ for LC mode $a$ with assistant
of a superconducting qubit. Here $|0\rangle$ and $|1\rangle$ are vacuum and Fock state with $1$ photon of mode $a$.
Then, we turn on the beam-splitter Hamiltonian (\ref{eq:Heff}) between $a$ and $a_{\rm m }$.
After the interaction time $t= \pi/g$, the mechanical mode will be in the superposition state $|\phi_0\rangle$. Here
we suppose that the strong coupling condition $g > \bar{n}_{\rm m } \gamma, \kappa$ fulfills.  Therefore, the coherence of
the superposition state $|\phi_0\rangle$ can maintain during the state transfer.

The quantum state of the center-of-mass motion of a microorganism can also be teleported to another microorganism using a superconducting  circuit. Quantum teleportation based on superconducting circuits has been demonstrated recently \cite{Steffen2013}. We consider two remote microorganisms, which are
attached to two separate mechanical resonators integrated with LC resonators. They are connected by a superconducting circuit as demonstrated in Ref. \cite{Steffen2013}. They are initially cooled down to the motional
ground states. We use the ground state and the first Fock state $|1\rangle$ of both mechanical and LC resonators as the qubit
states. The mechanical mode $a_{m1}$ of the first microorganism and mechanical resonator
is prepared to a superposition state $|\psi_1\rangle = \alpha |0\rangle_{m1}+\beta |1\rangle_{m1}$, where $\alpha$ and $\beta$ are arbitrary and
fulfill $|\alpha|^2 + |\beta|^2=1$. The LC resonator modes $a_1$ and $a_2$ are prepared to the entangled state $(|0\rangle_1 |1\rangle_2+ |1\rangle_1 |0\rangle_2)/\sqrt{2}$, through quantum state transfer, or post selection \cite{Yin2015}. Then
by transferring the state  $a_2$ to the mechanical mode $a_{m2}$ of the second microorganism and mechanical oscillator, the LC mode $a_1$ entangles with mechanical mode $a_{m2}$. This entanglement can be used as a resource for  teleporting the state in mechanical mode $a_{m1}$ to the mechanical mode $a_{m2}$ \cite{Bennett1993}.

To do this task, we need to  perform  Bell measurements
on mode $a_1$ and $a_{m1}$, which can be accomplished by a CPHASE gate between $a_1$ and $a_{m1}$, and Hadamard gates
on $a_1$ and $a_{m1}$ \cite{Steffen2013}.
In order to realize the CPHASE gate, we tune the driving detuning ${\mathnormal{\Delta}}-\omega_{\rm m }= \delta \ll \omega_{\rm m }$. The effective
Hamiltonian between $a_1$ and $a_{m1}$ becomes
\begin{equation}\label{eq:Heff1}
  H_{\rm eff}'= \hbar \delta a_1^\dagger a_1 +\hbar g a^\dagger a_{\rm m } + \hbar g a a_{\rm m }^\dagger.
\end{equation}
In the limit $\delta \gg g$, we get an effective photon-phonon coupling Hamiltonian by adopting
second order perturbation method,
\begin{equation}\label{eq:Hpp}
  H_{\rm pp}= \hbar \frac{g^2}{\delta} a_1^\dagger a_1 a_{m1}^\dagger a_{m1}.
\end{equation}
By turning on the Hamiltonian (\ref{eq:Hpp}) for time $t=\pi \delta /g^2$, the CPHASE gate between $a_1$ and
$a_{m1}$ accomplishes. The state of the system becomes $|{\mathnormal{\Psi}}\rangle = (\alpha|0\rangle_{m1} |0\rangle_1 |1\rangle_{m2}
+ \alpha |0\rangle_{m1} |1\rangle_1 |0\rangle_{m2} + \beta |1\rangle_{m1} |0\rangle_1 |1\rangle_{m2} - \beta |1\rangle_{m1}
|1\rangle_1 |0\rangle_{m2})\sqrt{2}$. Then we perform two Hadamard gates on $a_1$ and $a_{m1}$ modes, and project measurements
on the basis $|0\rangle$ and $|1\rangle$. There are four different output $\{00,01,10,11\}$, which are one-to-one mapping
to the four Bell measurement basis. Based on the output, we can perform the specific local operation on the mode $a_{m2}$ and recover the
original state $\alpha |0\rangle + \beta |1\rangle$.

\section{Internal electron spins: entanglement, detection and teleportation}
The internal states of a microorganism can also be prepared in superposition states and entangled with the center-of-mass motion of the microorganism at millikelvin temperatures.  A good  internal state of a microorganism is the electron spin of a radical or transition metal ion in the microorganism. Radicals are produced during metabolism or by radiation damage. Some proteins in microorganisms contain transition metal ions, which also have non-zero electron spins can be used to create quantum superposition states.  The smallest amino acid  in proteins is glycine. The electron spin of a glycine radical $\textrm{NH}^+_3{\dot{\textrm{C}}}\textrm{HCOO}^-$  has a relaxation time $T_1 = 0.31 $ s and a phase coherent time $T_{\rm M} = 6~ \mu$s at 4.2 K \cite{Hoffmann1995}. As shown in Fig. 4 of Ref. \cite{Hoffmann1995}, the phase coherent time increases dramatically when the temperature decreases below 10~K. So it will be much longer at millikelvin temperatures. Moreover, universal dynamic decoupling can be used to increase the coherent time $T_M$ by several orders, eventually limited by the relaxation time $T_1$ \cite{Lange2010}. Thus we  expect the coherent time of the electron spin of a glycine radical to be much longer than 1 ms at millikevin temperatures. The electron spin of some other radicals or defects in a frozen microorganism may have even longer coherence time at millikelvin temperatures.

\begin{figure}[btp]
\setlength{\unitlength}{1cm}
\begin{picture}(7,7)
\put(0,0){\includegraphics[totalheight=7cm]{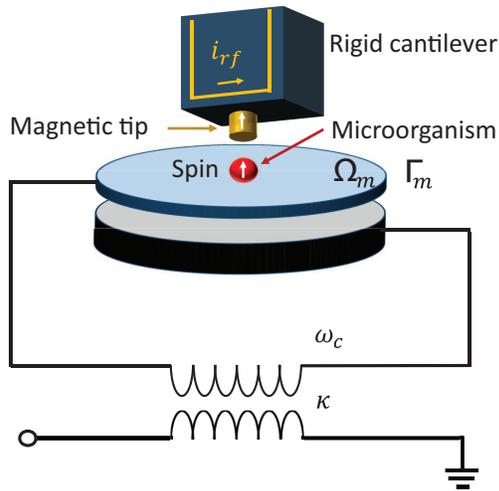}}
\end{picture}
\caption{(Color online) Scheme to couple the internal states of a microorganism to the center-of-mass motion of the microorganism with a magnetic gradient. A ferromagnetic tip  mounted on a rigid cantilever produces a strong magnetic gradient. A RF current $i_{\rm RF}$ passing through a superconducting microwire generates a $B_{\rm RF}$ that excites an electron  spin in the microorganism.  Different from a typical MRFM apparatus, the cantilever is rigid, while the substrate (membrane) of the microorganism is flexible to oscillate in this setup }
\label{fig:scheme2}
\end{figure}

As shown in Fig. \ref{fig:scheme2}, our scheme to couple the spin state and the center-of-mass motion of a microorganism with a magnetic tip is similar to the scheme used in MRFM. Recently, single electron spin detection with a MRFM  \cite{Rugar2004}, and nanoscale magnetic resonance imaging of the   $^1$H nuclear spin of tobacco mosaic viruses \cite{Degan2009} have been realized. A MRFM  using a superconducting quantum interference device (SQUID)-based cantilever detection at 30~mK has also been demonstrated \cite{Vinante2011}.

As shown in Fig. \ref{fig:scheme2}, in order to couple the internal spins states of a  microorganism to the center of mass motion of the
microorganism, a magnetic field gradient  is applied. Above the microorganism, there is a ferromagnetic tip mounted on a
rigid cantilever, which produces a magnetic field $\bf{B}$ with a large gradient. In a microorganism, there are usually more than one radical that
has unpaired electrons. Because the magnetic field is inhomogeneous, the energy splitting between electron spin states depends on the relative position between an electron and the
ferromagnetic tip. The Hamiltonian for $N$ unpaired electron spins reads
\begin{equation}\label{eq:electron}
  H_e= \sum_i^N \hbar g_{\rm s} \mu_{\rm B} \mathbf{S}_i (\vec{x}_i) \cdot \mathbf{B} (\vec{x}_i),
\end{equation}
where $g_{\rm s}\simeq 2$, $\mu_{\rm B}$ is Bohr magneton, $\vec{x}_i$ is the position of the $i$th electrons,
$\mathbf{S}_i$ is the spin operator for $i$th electrons spins. To make sure that the electron spin of interest is initially in the ground state at 10 mK, we suppose that the electron spin level spacings are larger than 500 MHz, which requires a magnetic field larger than 18 mT at the position of the radical. Because the finite size of the microorganism, the magnetic field at the position of membrane can be smaller than 10 mT, which is the critical magnetic field of aluminium. So the membrane can be made of aluminium \cite{Teufel11} if necessary. However,  superconducting materials (e.g. Nb) with a relative large critical magnetic field  will be better for making the  membrane and the RF wire \cite{Suh2012,Vinante2011}.

The oscillation of membrane induces a time-varying magnetic field on electrons in the microorganism. We define the
single phonon induced frequency shift $\lambda = g_{\rm s} \mu_{\rm B} |\mathbf{G}_{\rm m }| x_o'/\hbar$, where $x_0'$ is
the zero field fluctuation of microorganism, and $\mathbf{G}_{\rm m }
= \partial \mathbf{B}(\vec{x}_1)/ \partial \vec{x}_1$. Here $x_0'$ is different from $x_0$ of
membrane. Usually, $x_0'$ is two times larger than $x_0$ when the microorganism is at the center of the membrane. Here we assume that the magnetic gradient is (un)parallel to both the magnetic field $\mathbf{B} (\vec{x}_1)$ and the mechanical oscillation.
The $z$ axis is defined along the direction of $\mathbf{B} (\vec{x}_1)$.
We apply a microwave driving  with frequency $\omega_{\rm d}'$, which is close to the electron $1$'s level spacing $\omega_1=g\mu_{\rm B} B(\vec{x}_1)$.
The Rabi frequency is ${\mathnormal{\Omega}}_{\rm d }'$, which is tuned to near the mechanical mode frequency $\omega_{\rm m }$.
In all electrons, the level spacing $\omega_2=g\mu_{\rm B} B(\vec{x}_2)$ of the $2$nd
electron is closest to the $1$st one. Under the condition that $g_{\rm s}\mu_B |\mathbf{B}(\vec{x}_1)- \mathbf{B} (\vec{x}_2) |\gg
|{\mathnormal{\Omega}}|$, we have $|{\mathnormal{\Omega}} |\ll |\omega_1 -\omega_2|$. Therefore, only electron $1$ is
driven by the microwave. We can neglect the effects of all other free electrons in microorganism.
The Hamiltonian contains electron $1$ and the motion of the membrane and the  microorganism reads \cite{Rabl2009}
\begin{equation}\label{eq:EleMem}
  H_{\rm eM}= \hbar \omega_{\rm m } a_{\rm m }^\dagger a_{\rm m }+ \frac{{\mathnormal{\Delta}}_e}{2} \hbar \sigma_z + \frac{{\mathnormal{\Omega}}_{\rm d }'}{2} \hbar
  \sigma_x + \frac{1}{2} \hbar \lambda (a_{\rm m }+ a_{\rm m }^\dagger) \sigma_z,
\end{equation}
where ${\mathnormal{\Delta}}_e= \omega_{\rm d }'-\omega_1$. We rotate the electron spin axis to diagonalize it, which has the
effective level splitting $\omega_{\rm eff}= \sqrt{{\mathnormal{\Delta}}^2_e+ {\mathnormal{\Omega}}'^2_{\rm d }}$.

We first need to identify the position and  frequency of an electron spin  in the microorganism. We can search it by scanning the magnetic tip above the microorganism \cite{Rugar2004}.
The microwaves driving the LC resonator and the electron spins are turned on with Rabi frequencies ${\mathnormal{\Omega}}_{\rm d }$ and
${\mathnormal{\Omega}}_{\rm d }'$, and frequencies $\omega_{\rm d }$ and $\omega_{\rm d }'$. Once the electron spin resonance (ESR) reaches
$\omega_{\rm eff}=\omega_{\rm m }$, the excitation of the electron spin could be efficiently  transferred
to the LC mode $a$, mediated with the mechanical mode $a_{\rm m }$,  and decay finally. We will observe the ESR signal
on the output spectrum of the LC mode. Then we want to tune the ${\mathnormal{\Delta}}_e$ to be zero and maximize the
spin-phonon coupling strength.  First, we scan the frequency of the microwave on the electron spins, under the specific Rabi
frequency ${\mathnormal{\Omega}}_{\rm d }'$.  The ESR peaks appear in pair with the center frequency $\omega_1$. Then we tune
the frequency of the microwave drive on electron  $\omega_{\rm d }'=\omega_1$, the ESR should have only one peak. In this way, we find the resonant frequency $\omega_1$ of the
electron $1$. The magnetic field $\mathbf{B}$ near the electron can be got from Eq. (\ref{eq:electron}).
Then we can identify the relative location between the electron $1$ and the magnetic tip. This setup can also detect nuclear spins with lower sensitivity, which has broad applications \cite{Fisher2015,Degan2009, Cai2013}.

Let's suppose that ${\mathnormal{\Delta}}_e$ has been tuned to be zero, and denote the operators $\sigma_\pm = \sigma_z \pm {\rm i}\sigma_y$,
The qubit is defined on the eigenstates of $\sigma_x$.
In the limit $\lambda \ll \omega_{\rm m }, |{\mathnormal{\Omega}}_{\rm d }'|$,  we set ${\mathnormal{\Omega}}_{\rm d }' = \omega_{\rm m }$ and
the effective interaction Hamiltonian between electrons and the motion of microorganism is
 \begin{equation}\label{eq:HI}
   H_{\rm I}= \hbar \lambda \sigma_+ a_{\rm m } + h.c..
 \end{equation}
Here the rotating wave approximation is used. We may also set ${\mathnormal{\Omega}}_{\rm d }'= - \omega_{\rm m }$,
and the effective Hamiltonian has the form
 \begin{equation}\label{eq:HII}
   H_{\rm I}'= \hbar \lambda \sigma_+ a_{\rm m }^\dagger + h.c..
 \end{equation}

 If we use the parameters in Ref. \cite{Teufel11}, $\omega_{\rm m }=2\pi \times 10$ MHz, $M_{\rm mem}= 48$ pg,
 we have $x_0= 4.2 \times 10^{-15}$ m, and $x_0' \simeq 2x_0= 8.4 \times
 10^{-15}$ m. Under the magnetic gradient $|\textbf{G}_{\rm m }|=10^7$~T/m, the single phonon
 induced frequency shift $\lambda = g_{\rm s} \mu_{\rm B} |\textbf{G}_{\rm m }| x_o'/\hbar= 14.8$~kHz.
The mechanical decay rate $\gamma_{\rm m }= 2\pi \times 32$ Hz. The effective mechanical decay under temperature
$10$ mK is $\bar{n}_{\rm m } \gamma_{\rm m } \simeq 20\times 2\pi \times 32$~Hz $=4.0$~kHz, which is much less than $\lambda$. Recently, a mechanical decay rate as low as $\gamma_{\rm m }= 2\pi \times 9.2$ Hz was realized in a similar electromechanical membrane oscillator \cite{Lecocq2015}, which will be even better.
The electron spin decay and dephase  can be both less than kHz \cite{Hoffmann1995,Lange2010}, which is also much less than $\lambda$.
Therefore, strong coupling condition is fulfilled.  We can generate entangled state and transfer quantum
states between electron spin $1$ and the mechanical mode $a_{\rm m }$ with either Hamiltonian (\ref{eq:HI})
or (\ref{eq:HII}) \cite{Yin2013,Yin2015b}. Thus this setup not only can detect the existence of single electron spins like conventional MRFM  \cite{Rugar2004}, but  also can coherently manipulate and detect the quantum states of electron spins. It enables some isolated electron spins that could not be read out with optical or electrical methods to be used as quantum memory for quantum information.
To further increase the spatial separation of the superposition state of a microorganism, one can attach the microorganism to a magnetically levitated superconducting microsphere \cite{Romero-Isart2012b,Cirio2012,Geim1997} instead of a fixed membrane in future.

The internal electron spin state of a microorganism can also be teleported  to another microorganism using superconducting circuits \cite{Steffen2013}. We can first transfer the internal electron spin state of microorganism 1 to its mechanical state with either Hamiltonian (\ref{eq:HI}) or (\ref{eq:HII}) \cite{Yin2013,Yin2015b}. We then teleport it to the  mechanical state of the remote microorganism 2 as discussed in Sec. 3. Finally, we transfer the mechanical state of microorganism 2 to its internal electron spin state. In this way, we achieve the quantum teleportation between  internal electron spin states of two  microorganisms. In future,  this method can be extended to entangle and teleport multiple  degrees of freedom \cite{Heilmann2015,Sheng2010,Wang2015} of a living organism at the same time. Since internal states of an organism contain  information, our proposal also provides an experimental scheme for teleporting  information or memories between two remote organisms.

\section{Conclusions}
In summary, we propose a straightforward method  to create quantum superposition states of a  living microorganism by putting  a small cryopreserved bacterium on top of an electromechanical oscillator. The internal state of a microorganism, such as the electron spin of a glycine radical, can be entangled with its center-of-mass motion. Beyond ``Schr{\"o}dinger's cat'' thought experiment, we also propose schemes to teleport  the center-of-mass motion  and internal states of one microorganism to another remote microorganism.  Different from using simple inorganic samples, the use of a microorganism will allow this system to study defects and structures of proteins and other biologically important molecules. More importantly, our scheme can be used to study quantum wave function collapse due to biochemical reactions in future. At cryogenic temperatures, we can  enable certain photochemical reactions  with photons. For example,  the wavelike energy transfer through quantum coherence in photosynthetic systems was first observed at a cryogenic temperature \cite{Brixner2005}. We can put  a prochlorococcus, which is a photosynthetic organism, on top of an electromechanical oscillator and prepare its center-of-mass motion in a superposition state. We can then use one or a few photons to trigger the photosynthetic reaction of the prochlorococcus to study the effect of the internal biochemical reactions  on the collapse of the superposition state of the center-of-mass motion of a microorganism.

\vspace{0.5cm}
\textbf{Acknowledgments}
TL would like to thank the support from Purdue University and helpful discussions with G. Csathy, F. Robicheaux, C. Greene, and V. Shalaev.
ZQY is funded by the National Basic Research Program of China (2011CBA00300 and 2011CBA00302), the National Natural Science Foundation of China (11105136, 11474177 and 61435007). ZQY would like to thank the
useful discussion with Lei Zhang.

\vspace{0.5cm}
\textbf{Conflict of interest}
The authors declare that they have no conflict of interest.

\end{document}